\documentclass[aps,pra,twocolumn,superscriptaddress,floatfix,longbibliography]{revtex4-2}

\usepackage[utf8]{inputenc}
\usepackage[T1]{fontenc}
\usepackage{bm}
\usepackage[english]{babel}	
\usepackage{breakcites}
\usepackage{lmodern}
\usepackage[normalem]{ulem}

\usepackage{ae,aecompl}
\usepackage{subfigure}
\usepackage{latexsym,stmaryrd}
\usepackage{graphicx}
\usepackage[hidelinks,linkcolor=blue,colorlinks=true,urlcolor=blue,citecolor=blue]{hyperref}
\usepackage{braket}
\usepackage{amsmath}
\usepackage{upgreek}
\usepackage{comment}
\usepackage{url}
\usepackage{amssymb}

    \newcommand{\ac}{a}
\newcommand{\ncav}{\overline{n}_\mathrm{cav}}
\newcommand{\ain}{s_\mathrm{in}}

\newcommand{\Wmod}{\Omega_\mathrm{VNA}}
\newcommand{\pin}{P_\mathrm{in}}
\newcommand{\go}{g_\mathrm{o,1}}
\newcommand{\ggo}{g_\mathrm{o,2}}

\usepackage{lineno}
\usepackage{cleveref}
\usepackage{graphicx,amsmath}
\usepackage{siunitx}

\begin{document}

\title{Multimode optomechanics with a two-dimensional optomechanical crystal}

\author{Guilhem Madiot}
\affiliation{Catalan Institute of Nanoscience and Nanotechnology (ICN2), CSIC and BIST, Campus UAB, Bellaterra, 08193 Barcelona, Spain}

\author{Marcus Albrechtsen}
\affiliation{Department of Electrical and Photonics Engineering, DTU Electro, Technical University of Denmark, Building 343, DK-2800 Kgs.\ Lyngby, Denmark}

\author{Clivia M. Sotomayor-Torres}
\affiliation{Catalan Institute of Nanoscience and Nanotechnology (ICN2), CSIC and BIST, Campus UAB, Bellaterra, 08193 Barcelona, Spain}
\affiliation{ICREA - Instituci\'o Catalana de Recerca i Estudis Avan\c{c}ats, 08010 Barcelona, Spain}

\author{S{\o}ren Stobbe}
\affiliation{Department of Electrical and Photonics Engineering, DTU Electro, Technical University of Denmark, Building 343, DK-2800 Kgs.\ Lyngby, Denmark}
\affiliation{NanoPhoton - Center for Nanophotonics, Technical University of Denmark, Ørsteds Plads 345A, DK-2800 Kgs.\ Lyngby, Denmark}

\author{Guillermo Arregui}
\thanks{guibra@dtu.dk}
\affiliation{Department of Electrical and Photonics Engineering, DTU Electro, Technical University of Denmark, Building 343, DK-2800 Kgs.\ Lyngby, Denmark}

\begin{abstract}
Chip-scale multimode optomechanical systems have unique benefits for sensing, metrology and quantum technologies relative to their single-mode counterparts. Slot-mode optomechanical crystals enable sideband resolution and large optomechanical couplings of a single optical cavity to two microwave-frequency mechanical modes. Still, previous implementations have been limited to nanobeam geometries, whose effective quantum cooperativity at ultralow temperatures is limited by their low thermal conductance. In this work, we design and experimentally demonstrate a two-dimensional mechanical-optical-mechanical (MOM) platform that dispersively couples a slow-light slot-guided photonic-crystal waveguide mode and two slow-sound $\sim 7$ GHz phononic wire modes localized in physically distinct regions. We first demonstrate optomechanical interactions in long waveguide sections, unveiling acoustic group velocities below 800 m/s, and then move on to mode-gap adiabatic heterostructure cavities with a tailored mechanical frequency difference. Through optomechanical spectroscopy, we demonstrate optical quality factors $Q \sim 10^5$, vacuum optomechanical coupling rates, $g_o/2\pi$, of 1.5 MHz and dynamical backaction effects beyond the single-mode picture. At larger power and adequate laser-cavity detuning, we demonstrate regenerative optomechanical oscillations involving a single mechanical mode, extending to both mechanical modes through modulation of the input laser drive at their frequency difference. This work constitutes an important advance towards engineering MOM systems with nearly degenerate mechanical modes as part of hybrid multipartite quantum systems.
\end{abstract}

\maketitle

\section{Introduction}

The study of the interaction between an electromagnetic and a mechanical resonator in a cavity-optomechanical system has led to scientific and technological advances in quantum physics, nonlinear optics, and condensed matter physics~\cite{aspelmeyer2014cavity, li2021cavity, meystre2013short}. However, the canonical theoretical description of the one-to-one interaction~\cite{aspelmeyer2014cavity} fails to describe certain phenomena observed in realistic devices which naturally host multiple optical and mechanical modes~\cite{nielseN_multimode_2017}, i.e., multimode optomechanical systems, which can cause quantum decoherence when the collective interaction is uncontrolled ~\cite{PhysRevA.82.013818,pluchar2023thermal}. In the usual experimental setting with a single laser drive, undriven optical modes are usually unimportant. However, the individual parametric optomechanical couplings of the various mechanical modes to the driven optical mode lead to an effective coupling between them. The case of two mechanical modes, i.e., mechanical-optical-mechanical (MOM) systems, has witnessed particular attention due to its potential impact in the quantum regime, e.g. to probe decoherence processes ~\cite{PhysRevA.88.063850,PhysRevA.97.063832, li2019quantum}, to introduce nonreciprocity~\cite{PhysRevLett.125.023603}, or to enhance robustness to thermal noise via mode squeezing ~\cite{huang2023controllable}. In addition, its linearized Hamiltonian description is particularly suited for the study of exceptional points~\cite{PhysRevLett.109.223601, xu2016topological, djorwe2019exceptional}, while strongly-driven MOM devices exhibit collective nonlinear dynamics, including mode competition~\cite{Zhang2018, PhysRevE.98.032201}, synchronization~\cite{PhysRevA.96.023805,sheng_self_organized_2020}, and bistability control~\cite{PhysRevLett.129.123603}. At the frontier between these two fields of study --- namely the linear and nonlinear regimes --- lies the so far largely unexplored paradigm of nonlinear non-Hermitian physics, where phenomena like topological mode transfer or unidirectional phonon emission could be realized based on high-frequency coherent phonon self-oscillations~\cite{Benzaouia2022,ji2022tracking,fischer2023controlling}.

The optically mediated coupling between mechanical modes in MOM systems scales inversely with their frequency difference and becomes symmetrical only when their respective vacuum optomechanical coupling rates to the common optical cavity mode, $\go$ and $\ggo$, are identical. This has fostered research on physical implementations that exhibit (nearly) degenerate mechanical modes, where the modes coherently mix into optomechanically dark and bright dressed states~\cite{lin_coherent_2010} and can transfer energy efficiently~\cite{Shkarin2014}. These include Fabry-Pérot microcavities with two identical membranes~\cite{gartner_integrated_2018}, membrane-in-the-middle cavities with symmetry-enforced (high-order) mechanical mode degeneracies~\cite{Shkarin2014}, or parallel evanescently coupled optical resonators. The latter category exploits the strong dependence of the frequencies of the resulting optical supermodes on the distance between the resonators and embraces double-disk microcavities~\cite{lin_mechanical_2009}, bilayer photonic-crystal cavities~\cite{roh_strong_2010}, and photonic-crystal zipper cavities~\cite{eichenfield_picogram-_2009}. Interestingly, the electromagnetic boundary conditions across material interfaces lead to a strong local field enhancement, i.e., a slot-mode effect, when such distances are deep sub-wavelength and the field polarization is adequate~\cite{albrechtsen_nanometer-scale_2022}. Such an effect is non-resonant; therefore, the formation of supermodes is not required. This has allowed integrated MOM devices with large optomechanical couplings by using triple nanobeam geometries separated by tenths of nanometers, each supporting one of the excitations~\cite{Grutter:15}. These quasi-one-dimensional (1D) MOM slot-mode optomechanical crystals (OMCs) can display microwave-frequency phononic-crystal cavity modes and reach the sideband-resolved regime, which has recently allowed the observation of mechanical exceptional points~\cite{wu_-chip_2023}. However, their geometries may be unsuitable to study emergent macroscopic quantum phenomena in millikelvin MOM systems because of inefficient thermalization~\cite{ren_two-dimensional_2020}. In the presence of residual absorption, the limited heat dissipation pathways of 1D OMCs lead to a phononic hot bath which can destroy the prepared quantum states. In addition, the poor stiffness of nanobeams makes them prone to surface-force-induced collapses during and after fabrication~\cite{buks_stiction_2001}, limiting how narrow the gap between them can be, i.e., how large $g_{\text{o}}$ can be~\cite{Grutter:15, midolo_noems_2018}, and potentially require stress-release management~\cite{camacho_characterization_2009, Grutter:15}. 

Two-dimensional (2D) optomechanical structures can sustain even larger optical quality factors~\cite{safavi-naeini_optomechanics_2010, winger_chip-scale_2011}, large $g_{\text{o}}$ to long-lived hypersonic mechanical modes~\cite{ren_two-dimensional_2020,kersul_silicon_2023}, and have the additional benefits of enhanced heat dissipation~\cite{haret_extremely_2009,ren_two-dimensional_2020} and convenient stiffness. Nonetheless, to the best of our knowledge, no experimental work has focused so far on MOM experiments in 2D OMCs. Here we propose a novel waveguide and cavity optomechanics platform that enables the coupling of a slot-guided optical mode to two independent, nearly degenerate, microwave-frequency mechanical modes. By building mode-gap adiabatic heterostructure cavities~\cite{song_ultra-high-q_2005}, we demonstrate a sideband-resolved system with $g_{\text{o}}$ as high as \SI{1.5}{MHz} between C-band telecom photons in a cavity with a $Q \sim 10^5$ and two $\sim\SI{7}{GHz}$ acoustic resonators. Passive control over the frequency difference of the latter two via a geometrical parameter is achieved, which may enable tailored MOM systems adaptable to specific experimental requirements. By performing wavelength and power-dependent optomechanical spectroscopy of a device with mechanical modes only differing in frequency by 6 MHz, we provide evidence of multimode dynamical backaction in good agreement with the linearized optomechanical equations of motion of a MOM system. Finally, we demonstrate simultaneous self-oscillatory dynamics of two mechanical resonators stimulated by an intermodulation tone, consistent with recent demonstrations using two mechanical modes of a single nanobeam OMCC~~\cite{mercade_floquet_2021}. 

\section{Slot-mode multimode optomechanical crystal waveguides}

\begin{figure}[t]
    \centering
    \includegraphics[width=0.95\columnwidth]{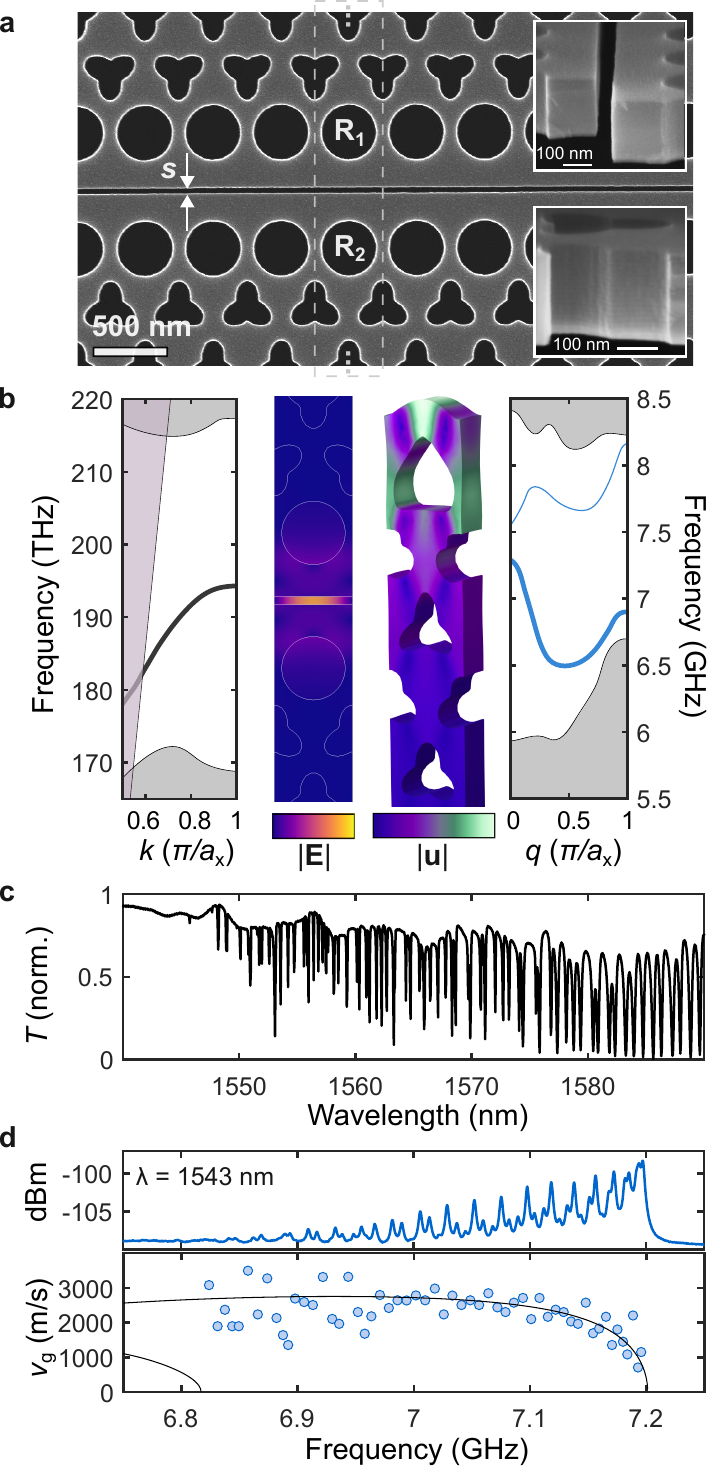}
    \caption{\textbf{Multimode optomechanical crystal waveguides (OMCWs).} (a) Scanning electron micrograph of an OMCW that couples two phononic waveguide modes --- one on each side of the air slot --- to a slot-guided optical mode. Insets: cross-section of the slot (top) and a cleaved shamrock. (b) Optical (left) and mechanical (right) dispersion diagrams of the structure in (a). The Bloch modes of interest are represented at $k = \pi/a_{\text{x}}$ and $q=0$. (c) Optical transmission spectrum of a mirror-terminated OMCW ($L$ = 350$a_x$) through a tapered fiber loop. (d) Radio-frequency (RF) spectrum measured with a laser drive on an optical mode at $\lambda = \SI{1543}{nm}$, and comparison of the simulated (line) and experimentally reconstructed (blue dots) acoustic group velocity.}
    \label{fig:MOMwaveguide}
\end{figure}

The geometry of the multimode OMC waveguides (OMCWs) we explore here consists of a line defect waveguide composed of two rows of circular holes with (optionally) different radii $R_1$ and $R_2$, a slot of width $s$, and two triangular lattices of shamrock-shaped~\cite{sollner_deterministic_2016,arregui_all-optical_2018} holes around them, with the shamrocks facing each other (Fig.~\ref{fig:MOMwaveguide}(a)). The structures are fabricated in \SI{220}{\nm} silicon-on-insulator (SOI) and the \SI{2}{\micro\meter} buried oxide is undercut to release the suspended structures with hydrofluoric vapour-phase etching. The patterns are defined in chemically semi-amplified resist with electron-beam lithography~\cite{albrechtsen_nanometer-scale_2022} and etched into the silicon using dry-etching based on a modified CORE process~\cite{nguyen_core_2020, madiot_optomechanical_2023}. The fabrication process is tailored to yield an excellent design-to-realized pattern fidelity, including smooth and vertical sidewalls (Fig.~\ref{fig:MOMwaveguide}(a), insets) ~\cite{albrechtsen_phdThesis_2023}. The presence of the slot mechanically decouples the two membrane sides, making the system a MOM optomechanical waveguide with a geometry-controlled mechanical frequency difference provided by $\Delta R = R_1 - R_2$. The employed shamrock crystal enables large near-infrared electromagnetic~\cite{arregui_all-optical_2018} and GHz mechanical band gaps~\cite{florez_engineering_2022}, and a similar geometry has been exploited for the in-situ generation of coherent acoustic phonons using Anderson-localized optical modes resulting from residual roughness in the etched sidewalls~\cite{madiot_optomechanical_2023}. However, the quality factors, $Q$, observed in Ref.~\cite{madiot_optomechanical_2023} were limited due to a multimode optical dispersion of the slot-guided mode and much below those measured in a slot photonic-crystal waveguide solely based on circular holes~\cite{arregui_cavity_2023}. The geometry we propose leverages the best features of both shamrock-shaped and circular holes as it exhibits single-moded dispersions with zero group velocity, i.e., simultaneous slow light and sound, at the mechanical and optical Bloch wavevectors where forward-type intra-modal Brillouin interactions are phase-matched~\cite{safavi-naeini_design_2010,kolvik_clamped_2023}, respectively at $q = 0$ and $k = \pi/a_{\text{x}}$ (Fig.~\ref{fig:MOMwaveguide}(b)). 
In addition, the vector parities of the Bloch optical ($y$- and $z$-symmetric $\mathbf{E}$ field) and mechanical ($z$-symmetric) modes make them optomechanically bright. The electric field amplitude of the optical mode, $|\mathbf{E}(\mathbf{r})|$, and the displacement amplitude, $|\mathbf{u}(\mathbf{r})|$, and deformation profile of the mechanical mode of interest are shown in Fig.~\ref{fig:MOMwaveguide}(b). The former exhibits subwavelength light confinement in the etched air slot, making the band edge frequency very sensitive to slot-width variations~\cite{Grutter:15, albrechtsen_nanometer-scale_2022}, while the latter is reminiscent of the in-plane breathing mode of a nanobeam waveguide, where one of the lateral free boundary conditions is replaced by a full-gap phononic crystal~\cite{patel_single-mode_2018}. The independent breathing motion of the two phononic waveguides (the mechanical mode is only represented for the bottom mechanical waveguide in Fig.~\ref{fig:MOMwaveguide}(b)) strongly modulates the slot width, which leads to large unit-cell vacuum optomechanical coupling rates, $g_{\text{o,cell}}/2\pi$, of 4.8 MHz, between the optical and the two mechanical Bloch modes. Departure from mechanical-mode degeneracy when the radii difference is $\Delta R \neq 0$ has a negligible effect on the respective values of $g_{\text{o,cell}}/2\pi$ (see Supplementary Section 1).

We probe the optical and mechanical properties of the OMCWs using a tunable diode laser connected to an optical fiber circuit that leads to a tapered fiber loop placed in contact with the slot waveguide. We terminate long OMCWs with short 32 unit cells waveguide segments within which the horizontal pitch is expanded from $a_{\text{x,1}}$ = 484 nm to $a_{\text{x,2}}$ = 510 nm. These sections behave simultaneously as optical and acoustic mirrors forming standing waves in the central waveguide region~\cite{madiot_optomechanical_2023} (see Supplementary Section 1 for dispersion diagrams as a function of $a_{\text{x}}$). Sharp spectral dips in the transmitted optical signal (Fig.~\ref{fig:MOMwaveguide}(c)) evidence evanescent coupling to resonant optical modes of the waveguide. We observe three spectral regions with distinct features: First, a region above $\lambda \sim \SI{1580}{nm}$ made of Fabry-Pérot optical modes with large on-resonance coupling fraction and a free spectral range (FSR) determined by the group index, $n_{\text{g}}$, and the length, $L$, of the waveguide. Second, the wavelength region $\SI{1560}{nm} < \lambda < \SI{1580}{nm}$, within which the coupling fraction is also large, but the resonant wavelengths are seemingly random. We attribute these resonances also to Fabry-Pérot modes whose wavelengths are perturbed by the presence and exact position of the loop (see next section for a discussion of the dispersive effects of the loop on a single resonant optical mode). Third, a region close to the band edge ($\lambda > \SI{1560}{nm}$) where strong slow-light-induced backscattering from sidewall roughness leads to Anderson-localized modes at random spectro-spatial locations, leading to a random FSR and strong mode-to-mode fluctuations in the coupling fraction~\cite{arregui_cavity_2023,madiot_optomechanical_2023}. 

We characterize the mechanical properties of the OMCWs by moderate power and blue-detuned driving of optical modes in the first and second regions. We detect the thermal motion of the Fabry-Pérot standing mechanical modes using a fast photoreceiver and an electronic spectrum analyzer. A characteristic radio-frequency (RF) spectrum for an Anderson-localized optical mode at $\lambda = \SI{1543}{nm}$ is shown in Fig.~\ref{fig:MOMwaveguide}(d). For the particular case shown ($\Delta R$ = 0), no degeneracy-lifting between the two independent phononic waveguide modes is observed. The spectrum is composed of many overlapping and regularly spaced Lorentzian-shaped mechanical resonances with estimated linewidths, $\Gamma$, in the range 2--5 MHz and a transduction amplitude envelope that grows as the mechanical band edge at \SI{7.2}{GHz} is approached. We attribute this overall shape to a relaxed sinc-like phase-matching condition between an optical mode with $k$-components dominated by $k = \pi/a_{\text{x,1}}$ and mechanical modes in the vicinity of $q$ = 0 for forward-type Brillouin scattering interactions in a finite waveguide~\cite{kolvik_clamped_2023}. This remains true even if the optical mode employed is Anderson-localized because its $k$-space representation is often centered around the wavevector of the underlying waveguide at that frequency~\cite{savona_electromagnetic_2011}, i.e. $k \sim \pi/a_{\text{x,1}}$. While Fig.~\ref{fig:MOMwaveguide}(d) only exhibits peaks within the single-mode regime, RF spectra obtained by driving other optical modes also reveal peaks in the multi-mode propagation regime, in which the FSR is ill-defined due to inter-modal mechanical mixing. Using five additional RF spectra (see Supplementary Section 3), we reconstruct the acoustic group velocity, $v_{\text{g}}$, of the top and bottom phononic waveguides for $\Delta R$ = 0. The reconstructed $v_{\text{g}}$ is shown at the bottom of Fig.~\ref{fig:MOMwaveguide}(d) and compared to the simulated $v_{\text{g}}$. The simulation curve has been rigidly offset by only $-\SI{90}{MHz}$ ($\sim\SI{1}{\percent}$ of $\Omega$) to account for potential systematic errors on the SEM-extracted contour of the geometric features, illustrating the good agreement between simulations and measurements. We measure slow propagation of $\sim$ 7 GHz acoustic waves down to a group velocity below 800 m/s, constituting a 7-fold reduction relative to the transverse speed of sound in bulk silicon.

\section{Slot-mode multimode optomechanical crystal cavities}

\begin{figure*}[t!]
    \centering
    \includegraphics[scale=0.65]{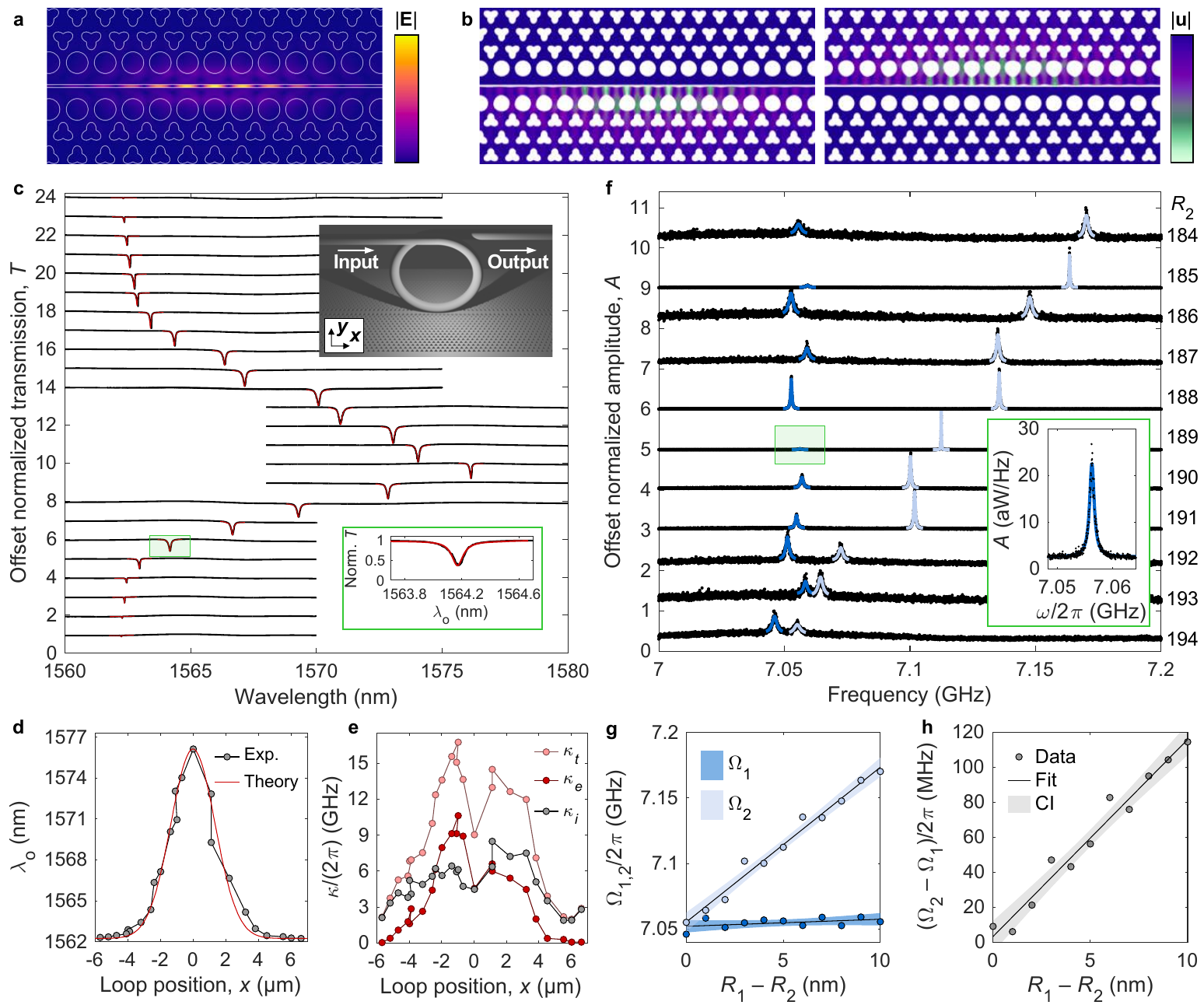}
    \caption{\textbf{Optomechanical spectroscopy of a 2D slot-mode mechanical-optical-mechanical (MOM) system.} (a) Electric field amplitude, $|\mathbf{E}|$, of the optical cavity mode and (b) displacement amplitude, $|\mathbf{u}|$, and deformation profiles of the two mechanical cavity modes. (c) Normalized transmission spectrum at different positions along the waveguide axis, $x$. The spectra are offset for clarity and the red curves are Lorentzian fits to the resonances. Insets: Schematic of the configuration with the loop on top of the cavity (top) and a close-up of a normalized spectrum (bottom). (d) Resonant wavelength as a function of the loop position, extracted from the fits in (a) and through a convolution of the calculated electric field intensity, $|\mathbf{E}|^2$, with a Gaussian ($\sigma=\SI{0.99}{\micro\m}$). (e) Total, external, and intrinsic decay rates as a function of the loop position. (f) Normalized radiofrequency (RF) spectra measured for devices with decreasing circular hole radius $R_2$. The spectra are offset for clarity. Blue solid lines are Lorentzian fits to the individual peaks. The inset shows a characteristic power spectral density of and fit to a single mechanical peak. (g) Extracted mechanical frequencies of the two plates (dots), linear fit (solid line) and confidence interval of the fit (shaded background). (h) Same as (e) for their frequency difference.}
    \label{fig:MOMcavities}
\end{figure*}

The efficient coupling of a multitude of Fabry-Pérot mechanical modes and Anderson-localized optical modes in the mirror-enclosed 2D OMCWs presented above provides an interesting platform to explore collective effects in multimode optomechanics, such as light squeezing~\cite{nielseN_multimode_2017}. However, some of the strongest optomechanical effects beyond the canonical one-to-one interaction occur in the MOM configuration. To explore such a setting, we engineer a MOM OMC cavity (OMCC) that couples two mechanical resonators and a high-$Q$ optical mode based on the waveguide modes of Fig.~\ref{fig:MOMwaveguide} and the respective partial band gaps above their band-edge frequencies. We adiabatically tune the horizontal pitch, $a_{\text{x}}$, along the waveguide axis from a central defect unit cell ($a_{\text{x,1}}$ = 484 nm) to mirror unit cells ($a_{\text{x,2}}$ = 510 nm) on both sides. The full defect region is formed by $N_{\text{c}}$ = 15 unit cells, and additional invariant sections made of $N_{\text{m}}$ = 32 mirror unit cells are included at the edges of the defect to prevent in-plane losses. Fig.~\ref{fig:MOMcavities}(a) shows the amplitude of the electric field, $|\mathbf{E}|$, of the cavity mode for $\Delta R$ = 0, whose theoretical resonant wavelength and quality factor are $\lambda_{\text{o}}$ = 1556 nm and $Q_{\text{i}}$ = 1.05 $\times 10^8$. The deformation profile and displacement amplitude, $|\mathbf{u}|$, of the two mechanical modes are shown in Fig.~\ref{fig:MOMcavities}(b), and both have mechanical frequency $\Omega_{1,2}/2\pi$ = 7.15 GHz in the degenerate case. The frequency difference between the two mechanically decoupled mechanical cavity modes is controlled by decreasing $R_2$ relative to $R_1$, which in turn slightly redshifts the resonant wavelength of the optical cavity mode. More details on the cavity-optomechanical figures of merit are provided in Supplementary Section 2.

We characterize the optical and mechanical properties of the MOM OMCCs using the same fiber-loop evanescent technique as in Fig.~\ref{fig:MOMwaveguide}. To account for the strong perturbative effect of the fiber loop on the resonant wavelength and losses of the optical cavity mode and infer their unperturbed parameters, i.e., in the absence of the loop, we systematically study the optical response as a function of the loop position.
Figure~\ref{fig:MOMcavities}(c) shows the optical transmission spectra across the cavity resonance for 24 different loop positions, $x$, with the loop in contact with the sample and approximately aligned to the slot axis (Fig.~\ref{fig:MOMcavities}(c) top inset). The position is extracted via analysis of microscope images acquired with a $100\times$ objective imaging the probed structure from above. By moving the sample under the loop while in contact, the loop slides along the slot and changes its overlap with the optical cavity mode. We extract the resonant wavelength, $\lambda_{\text{o}}$, and the extrinsic, $\kappa_{\text{e}}/2\pi$, and intrinsic, $\kappa_{\text{i}}/2\pi$, decay rates of the cavity mode for each loop position by fitting the cavity resonance with a Lorentzian lineshape (Fig.~\ref{fig:MOMcavities}(c) bottom inset) and using that the on-resonance transmission is given by $T_{\text{0}}$ = $(1-\kappa_{\text{e}}/\kappa_{\text{t}})^2$, with $\kappa_{\text{t}} = \kappa_{\text{e}} + \kappa_{\text{i}}$. Figure~\ref{fig:MOMcavities}(d) shows the extracted $\lambda_{\text{o}}$ along with a theoretical prediction obtained from a convolution between the calculated $|\mathbf{E}|^2$ and a Gaussian envelope representing the loop~\cite{albrechtsen_nanometer-scale_2022}. The Gaussian has a standard deviation, $\sigma=\SI{0.99}{\micro\meter}$, identified with a least-mean-square optimization. When the loop is centered on top of the cavity, the dispersive perturbation is maximal, shifting the wavelength by as much as $\sim\SI{15}{nm}$, i.e., \SI{1}{\percent} of the cavity wavelength. We identify the center position, $x$ = 0, as the position causing the largest redshift of the resonance wavelength. 
The evolution of the total, extrinsic and intrinsic decay rates as a function of the loop position are shown in Fig.~\ref{fig:MOMcavities}(e). Several interesting properties can be observed. First, $\kappa_{\text{i}}$ changes considerably with the loop position, which implies that the loop not only loads the cavity but also adds additional undetected loss pathways. Second, we observe a decrease of both $\kappa_{\text{i}}$ and $\kappa_{\text{e}}$ when the loop is centered on the cavity. We hypothesize that this is due to the additional symmetry of this configuration. Third, when the center of the loop is far from the geometric cavity center, the behaviour of $\kappa_{\text{i}}$ may indicate that the value is still unconverged, contrary to the wavelength. This indicates that the employed technique may not allow the measurement of the unperturbed optical linewidth. Therefore, we identify the unperturbed cavity parameters to be $\lambda_{\text{o}}$ = 1562 nm and $\kappa_{\text{i}}/2\pi < \SI{2}{GHz}$ (for the $R_2 = \SI{193}{nm}$ here). 

\begin{figure*}[t]
    \centering
    \includegraphics[scale=0.9]{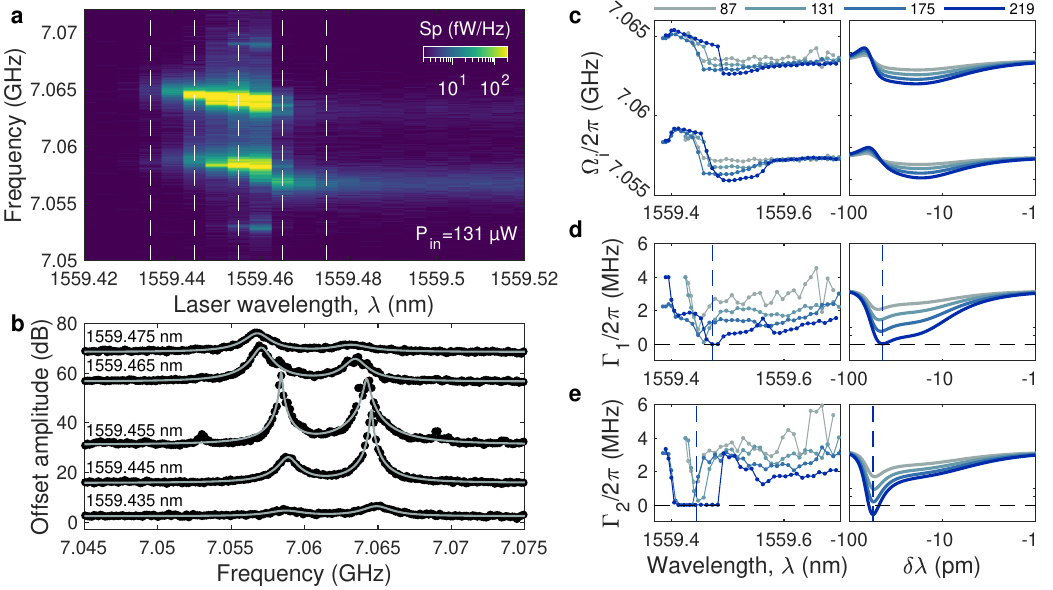}
    \caption{\textbf{Dynamical back-action in the nearly degenerate case.} (a) Measured radiofrequency (RF) spectra as a function of the laser wavelength $\lambda$ for a laser power $\pin=131$ $\mu$W. The dashed lines indicate the value of $\lambda$ for the spectra in (b). (b) Individual RF spectra (black dots) fitted with a double-Lorentzian function (grey solid lines) to extract the mechanical frequencies and damping rates. The curves are vertically offset for clarity. (c-e) Experimental (left) and theoretical (right) evolution of (c) the mechanical frequencies $\Omega_1$ and $\Omega_2$, and the mechanical damping rates (d) $\Gamma_1$ and (e) $\Gamma_2$. These parameters are plotted at four different powers (see legend with $\pin$ indicated in units of $\mu$W) and as a function of $\lambda$ or the laser-cavity detuning, $\delta\lambda$, respectively, for experiment and theory. The blue dashed lines indicate the wavelength (or detuning) at which the minimum of the damping rates occur for $\pin=219$ $\mu$W.}
    \label{fig:scan}
\end{figure*}

We systematically characterize the mechanical cavity modes for OMCCs with different values of $R_2$ as in Fig.~\ref{fig:MOMwaveguide}(d), with the fiber loop positioned to achieve a minimal perturbation and critical coupling to the optical cavity mode, i.e., $T_{\text{0}}=1/2$. 
The value of $R_2$ is nominally reduced in \SI{1}{nm} steps from $R_1=R_2=\SI{194}{nm}$ to $R_2=\SI{184}{nm}$. Figure~\ref{fig:MOMcavities}(f) shows the RF spectra with Lorentzian fits to both mechanical modes. Following the prediction of the finite-element simulations (see Supplementary Sections 1 and 2), we identify the lower (higher) frequency mechanical mode, fitted in dark (light) blue, as that of the top (bottom) membrane. This probably holds true except for the smaller values of $\Delta R$, where the effect of disorder-induced dispersion might overcome the as-designed deterministic frequency difference. We note that the relative amplitude of the transduced signals is determined by the exact loop position transverse to the slot axis, which determines to what extent the modes are dampened, and by the presence of dynamical backaction effects, which may occur preferentially for one of the modes at a fixed detuning~\cite{Zhang2018}. Despite this, the signal-to-noise ratio (SNR) remains sufficient to extract the central frequencies accurately. Figure~\ref{fig:MOMcavities}(g) shows the extracted frequencies as well as linear fits predicting $\Omega_1(\Delta R)/2\pi=7.052(2)\, \SI{}{GHz} + 0.5(4)\Delta R\ \SI{}{MHz\per\nm}$ and $\Omega_2(\Delta R)/2\pi=7.055(3)\, \SI{}{GHz} + 11.8(5)\Delta R\ \SI{}{MHz\per\nm}$. The non-zero slope found for $\Omega_1$ likely originates from short-range proximity effects~\cite{albrechtsen_nanometer-scale_2022,albrechtsen_phdThesis_2023}, i.e., the smaller value for $R_2$ leads to a smaller effective electron-beam dose on the other membrane side, which is not accounted for with standard long-range proximity effect correction~\cite{florez_engineering_2022}. Meanwhile, the slope of $\Omega_2$ agrees well with the simulation prediction (10.19(7) MHz/nm). Note that the mechanical frequencies are approximately 90 MHz lower than the simulated values, which is in very good agreement to the theory-experiment offset found for the acoustic waveguide band of Fig.~\ref{fig:MOMwaveguide}(d). Figure~\ref{fig:MOMcavities}(h) shows $\delta\Omega(\Delta R)/2\pi$ illustrating its linear dependence with an intercept $\delta\Omega(0)/2\pi=(3\pm4)\,\SI{}{MHz}$. This non-zero intercept corroborate that we consistently observe a separation of a few MHz between the nominally degenerate mechanical modes, which is caused by inherent fabrication imperfections. The values of $\kappa_{\text{t}}$ under the measurement conditions of Fig.~\ref{fig:MOMcavities}(f) are all in reasonable agreement with those reported in Fig.~\ref{fig:MOMcavities}(e), i.e., we observe no pronounced effect of $\Delta R$ on the optical $Q$ (as expected from simulations, see Supplementary Figure S3), and therefore the MOM OMCs we demonstrate are all in the sideband-resolved regime ($\kappa_{\text{t}}$<$\Omega_{1,2}$). In addition, the values of $\go/2\pi$ and $\ggo/2\pi$ are measured to be in the range of 1.2--1.5 MHz, which is in excellent agreement with simulations (see Supplementary Sections 2 and 4 for details on the simulated and measured $\go$ and $\ggo$). In the next section, we explore the implications of the reported cavity-optomechanical figures of merit and the MOM nature of the system on the detuning- and power-dependence of the dynamical backaction effects on a laser-driven device with nearly degenerate mechanical modes.

\section{Optomechanical spectroscopy of a mechanical-optical-mechanical system}

We consider a structure with $R_1=194$ nm and $R_2=193$ nm and place the loop in the position indicated previously, leading to loss rates of $\kappa_{\text{i}}/2\pi\approx3.5$ GHz and $\kappa_{\text{e}}/2\pi\approx0.76$ GHz, and a resonance wavelength (at low power) of $\lambda_{\text{o}}=1559.35$ nm. Using a laser power of $\pin=131$ $\mu$W, we step-scan the laser wavelength, $\lambda$, from the blue-detuned side of the optical resonance and measure the RF spectrum. In Fig.~\ref{fig:scan}(a), we show the resulting power spectral density as a colormap, highlighting the two mechanical resonances separated by approximately 6 MHz. To track the mechanical frequencies and linewidths as a function of $\lambda$, we fit each spectrum with a sum of two Lorentzians. Examples of the fitted spectra are presented in Fig.~\ref{fig:scan}(b), with experimental data extracted from the above map as indicated by the white dashed lines. The same procedure is applied for $\pin=$ 87, 131, 175 and 219 $\mu$W, and the extracted parameters are summarized in Fig.~\ref{fig:scan}(c-e) (left column). For increasing input power, the mechanical resonances experience an increasing displacement around their natural value (Fig.~\ref{fig:scan}(c)) due to optomechanical dynamical backaction. Meanwhile, the mechanical damping rates tend to decrease down to a minimum, as expected in the blue-detuned regime. For sufficiently high input power (here at $\pin=219$ $\mu$W), the damping rate saturates around $\sim10$ kHz, which indicates that the mode is self-oscillating~\cite{aspelmeyer2014cavity}. In Fig.~\ref{fig:scan}(c-e) (right column), we qualitatively compare the extracted parameters as a function of laser wavelength to the values predicted by the linearized optomechanical equations of motion of a MOM system~\cite{xu2016topological,ng2022intermodulation}. In addition to the optical parameters given above, the model uses $\Omega_{1,2}/2\pi = 7.061\text{ GHz }\pm 3.05$ MHz, $\Gamma_1/2\pi = \Gamma_2/2\pi = 3.2$ MHz, $\go/2\pi = 1.25$ MHz and $\ggo/2\pi = 1.5$ MHz. Details on the theoretical model are given in Supplementary Section 6. We note that the model ignores the observed residual-absorption-mediated thermal non-linearities, so the horizontal axis in the theoretical plots of Fig.~\ref{fig:scan}(c-e), which use the laser detuning $\delta\lambda=\lambda-\lambda_0$, cannot be directly mapped to the horizontal axis of the experimental plots as the true detuning scales non-linearly with the laser wavelength due to the thermo-optic drag, i.e. $\lambda_0 = f(\lambda)$. Nevertheless, the latter produces a nearly asymptotic decrease of $\delta\lambda$ towards zero (see Supplementary Section 5), followed by a sudden jump in the red-detuned regime ($\delta\lambda>0$). Therefore, theory and experiment can be qualitatively compared by using a logarithmic scale for the horizontal axis of the theoretical plots. We observe that the model captures the overall evolution of the mechanical frequencies and damping rates. In particular, the damping rates reach their minimum at different values of $\lambda$ (or $\delta\lambda$), as highlighted with the dashed lines for the case $\pin=219$ $\mu$W. This feature, which does not emerge in a model with two independent optomechanical oscillators (see Supplementary Section 5), suggests that the mechanical modes start hybridizing despite their non-identical mechanical frequencies. The same applies to any asymmetrical feature in the relative evolution of the parameters of both mechanical modes. We note that below $\lambda\approx1559.4$ nm, the SNR of the transduced mechanical resonances is very low, which prevents determining the frequencies and damping rates. Furthermode, above $\lambda\approx1559.7$, the properties of the modes stabilize, and their transduction slowly decreases until the laser exits the thermo-optic resonance.

\begin{figure}[ht]
    \centering
    \includegraphics[scale=0.95]{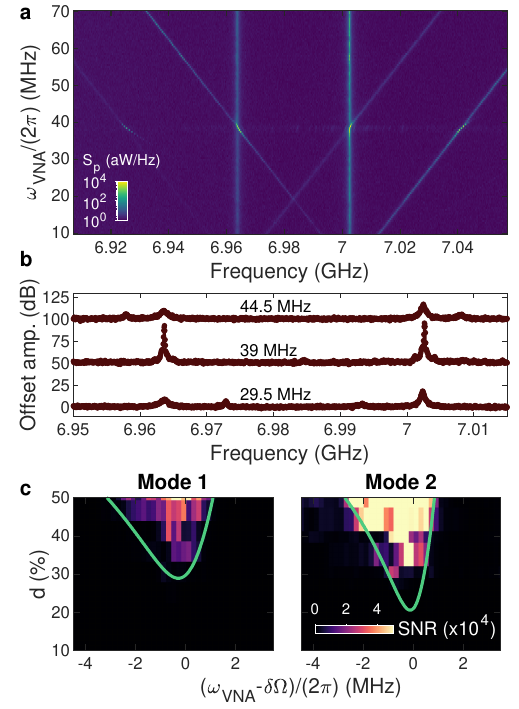}
    \caption{\textbf{Simultaneous Floquet mechanical lasing of two nearly degenerate mechanical modes.} (a) Measured RF spectra as a function of the modulation frequency, $\Wmod$. (b) Individual RF spectra extracted from (a) and vertically offset for clarity. (c) Measured RF peak amplitude as a function of the modulation depth and modulation frequency for mechanical modes 1 (left) and 2 (right). The green lines are the theoretical amplification threshold obtained using $\Delta=1.024\overline{\Omega}$ (with $\overline{\Omega}=6.984$ GHz) and the model in Ref.~\cite{mercade_floquet_2021}.}
    \label{fig:SMML}
\end{figure}

\section{Stimulated two-mode optomechanical amplification via intermodulation}

The simultaneous parametric amplification of two thermally excited mechanical oscillators coupled to a common optical mode is typically prevented by mode competition~\cite{Zhang2018} leading to anomalous cooling~\cite{kemiktarak_mode_2014}, except for the case of two mechanical resonators of disparate frequency~\cite{Grutter:15,ng2022intermodulation}. Nevertheless, such two-mode amplification can be stimulated by modulating the input laser intensity at the intermodal frequency~\cite{mercade_floquet_2021}. To demonstrate such a feature in the investigated MOM OMCC, we focus on a device with $R_1=194$ nm and $R_2=190$ nm, which results in a frequency difference of $\delta\Omega/2\pi\approx38$ MHz. We fix the optical loading conditions as before, with a blue-detuned driving at a power $\pin=90$ $\mu$W and considerably below self-oscillation threshold. We then apply direct intensity modulation of the laser with modulation depth $d$ and frequency $\Wmod$. We report in Fig.~\ref{fig:SMML}(a) the recorded RF spectrum for $d=30\%$ while step-scanning the modulation frequency. We observe the two mechanical modes at $\Omega_1/2\pi=6.965$ GHz and $\Omega_2/2\pi=7.003$ GHz with respective linewidths ---when the modulation is off--- $\Gamma_1/2\pi=2.417$ MHz and $\Gamma_2/2\pi=1.234$ MHz. Note that the mechanical frequencies are slightly lower than those reported in Fig.~\ref{fig:MOMcavities}(f-g-h) because the device we analyze here used a lithography mask with exposed (void) features shrunk by 5 nm uniformly. Each mechanical peak is surrounded by a pair of sidebands at a distance of $\pm\Wmod$. When the high-frequency (low-frequency) sideband of the peak at $\Omega_1$ ($\Omega_2$) crosses the mode at $\Omega_2$ ($\Omega_1$), i.e., when $\Wmod=\delta\Omega$, the amplitude of the mechanical peaks increases significantly, far beyond (by at least 20 dB) the value given by the sum of the thermal transduction and the sideband peak. In Fig.~\ref{fig:SMML}(b), we show three spectra extracted from the upper map, corresponding to $\Wmod<\delta\Omega$, in which case the two modes are thermally excited (bottom), $\Wmod\approx\delta\Omega$, where the modes amplify by nearly 40 dB (middle), and $\Wmod>\delta\Omega$, for which the amplitudes of the modes reduce down to that of the sub-threshold regime (top). 

In order to identify the dynamical range enabling stimulated multimode lasing to occur, we record the transduced amplitudes of both mechanical modes as a function of both modulation parameters $d$ and $\Wmod$. Figure~\ref{fig:SMML}(c) represents the SNR evaluated based on the mode amplitude in the absence of modulation, i.e. in its unaltered thermal regime. For each mode, we plot the lasing threshold calculated from the theory in Ref.~\cite{mercade_floquet_2021} using $\go/2\pi=\ggo/2\pi=1.5$ MHz, $\lambda_0=1579.895$ nm, $\kappa_{\text{i}}=2.074$ GHz, and $\kappa_{\text{e}}=0.664$ GHz. The theory is calculated using the optical detuning $\Delta$ as a fitting parameter, since the presence of thermo-optic effect prevents its independent determination. However, we note that the asymmetry of the amplification area about $\Wmod$ is finely determined by the laser detuning $\Delta$, and is symmetric when the optomechanical amplification is optimal, i.e. at $\Delta=\overline{\Omega}$, where $\overline{\Omega}=6.984$ GHz is the mean mechanical frequency. This allows an unambiguous fit using $\Delta=1.024\overline{\Omega}$, and leads to a qualitative agreement of the theory with the experimental findings. Note that the second mechanical mode reaches the lasing threshold at lower modulation depth because of its significantly lower damping rate ($\Gamma_2<\Gamma_1$).

\section{Conclusion and outlook}

In summary, we have demonstrated a 2D OMC design that can operate as a sideband-resolved multimode MOM cavity-optomechanical system with two $\sim7$ GHz mechanical modes at a close spectral distance, $\delta\Omega_{1,2}\ll \Omega_{1,2}$, and a high-$Q$ optical cavity ($Q \sim 10^5$). The measured optomechanical coupling rates reaching $g_o/2\pi \sim$ 1.5 MHz are amongst the largest values observed in OMCCs with microwave-frequency mechanical modes, enabling low-power self-oscillations of the individual mechanical modes and simultaneous lasing upon modulation of the laser drive at their frequency difference. The latter is shown to be controlled, starting from the nominally degenerate case, by slightly breaking a structural symmetry, which does not significantly degrade the aforementioned properties over the 120 MHz passive tuning range. By performing a wavelength and power-dependent spectroscopic analysis of a structure with nominally degenerate mechanical modes, we show that features of multimode optomechanical backaction are observed. The combination of stochastic and deterministic deviations of the fabricated structure from the nominal design introduces a built-in frequency difference that we observe to be lower-bounded to around 4 MHz, preventing stronger optically-induced hybridization and the exploration of exceptional points. We foresee that introducing thermal tuning elements on both sides of the waveguide~\cite{burgwal2023enhanced}, e.g., metallic bonding pads, will enable independent and low-cross-talk tuning of the mechanical frequencies, which will allow the proposed system to reach the degenerate case and explore the physics across the frequency-crossing. Given the investigated limitations imposed by the physical presence of the fiber loop on the optical and mechanical losses, we expect further improvement by incorporating butt-coupled~\cite{ren_two-dimensional_2020} or side-coupled input/output waveguides~\cite{sonar_towards_2022}. Another necessary avenue required to leverage the improved heat dissipation of the 2D OMCs is the passivation, via termination chemistry~\cite{borselli_measuring_2006} or encapsulation layers~\cite{borselli_surface_2007}, of the slot-sidewall defect states that lead to the thermo-optical bistability we observe and that generally prevents us from performing red-detuned optomechanical cooling. In addition, the prospects of self-assembling gaps way below the limits of top-down nanofabrication~\cite{babar_self-assembly_2023} may allow reaching $g_{\text{o}}$ rates about four times larger than the ones we already report. Finally, the strong transduction of Fabry-Pérot mechanical modes we observe on the mirror-terminated waveguides and the demonstrated slow-down of sound to $\sim$ 700 m/s, finds applications in injection-locked optomechanical oscillators~\cite{modica_slow_2020}, on-chip phonon networks for information processing~\cite{patel_single-mode_2018}, or phononic quantum memories~\cite{wallucks_quantum_2020}. Conversely, the sensitivity of slow sound to fabrication imperfection~\cite{garcia_optomechanical_2017} and large ensemble measurements on mirror-terminated waveguides may allow the observation of spectral features resulting from Anderson localization of hypersonic acoustic waves~\cite{arregui_anderson_2019}.

\begin{acknowledgements}
The authors aknowledge Karl Pelka for useful discussion. G.M.\ and C.M.S.T.\ acknowledge the support from the project LEIT funded by the European Research Council, H2020 Grant Agreement No.\ 885689.\ ICN2 is supported by the Severo Ochoa program from the Spanish Research Agency (AEI, Grant No.\ SEV--2017-0706) and by the CERCA Programme/Generalitat de Catalunya. M.A.\ and S.S.\ gratefully acknowledge funding from the Villum Foundation Young Investigator Program (Grant No.\ 13170). S.S.\ additionally acknowledges funding from the Danish National Research Foundation (Grant No.\ DNRF147--NanoPhoton), the Innovation Fund Denmark (Grant No.\ 0175-00022--NEXUS), the Independent Research Fund Denmark (Grant No.\ 0135-00315--VAFL), and the European Research Council (Grant. No. 101045396 -- SPOTLIGHT). G.A. acknowledges financial support from the European Union’s Horizon 2021 research and innovation programme under the Marie Skłodowska-Curie Action (Grant No. 101067606 - TOPEX).
\end{acknowledgements}

\clearpage
\appendix

\section{Two-dimensional optomechanical crystal waveguides: band structures}

\begin{figure}[ht]
\centering
\includegraphics[width=\linewidth]{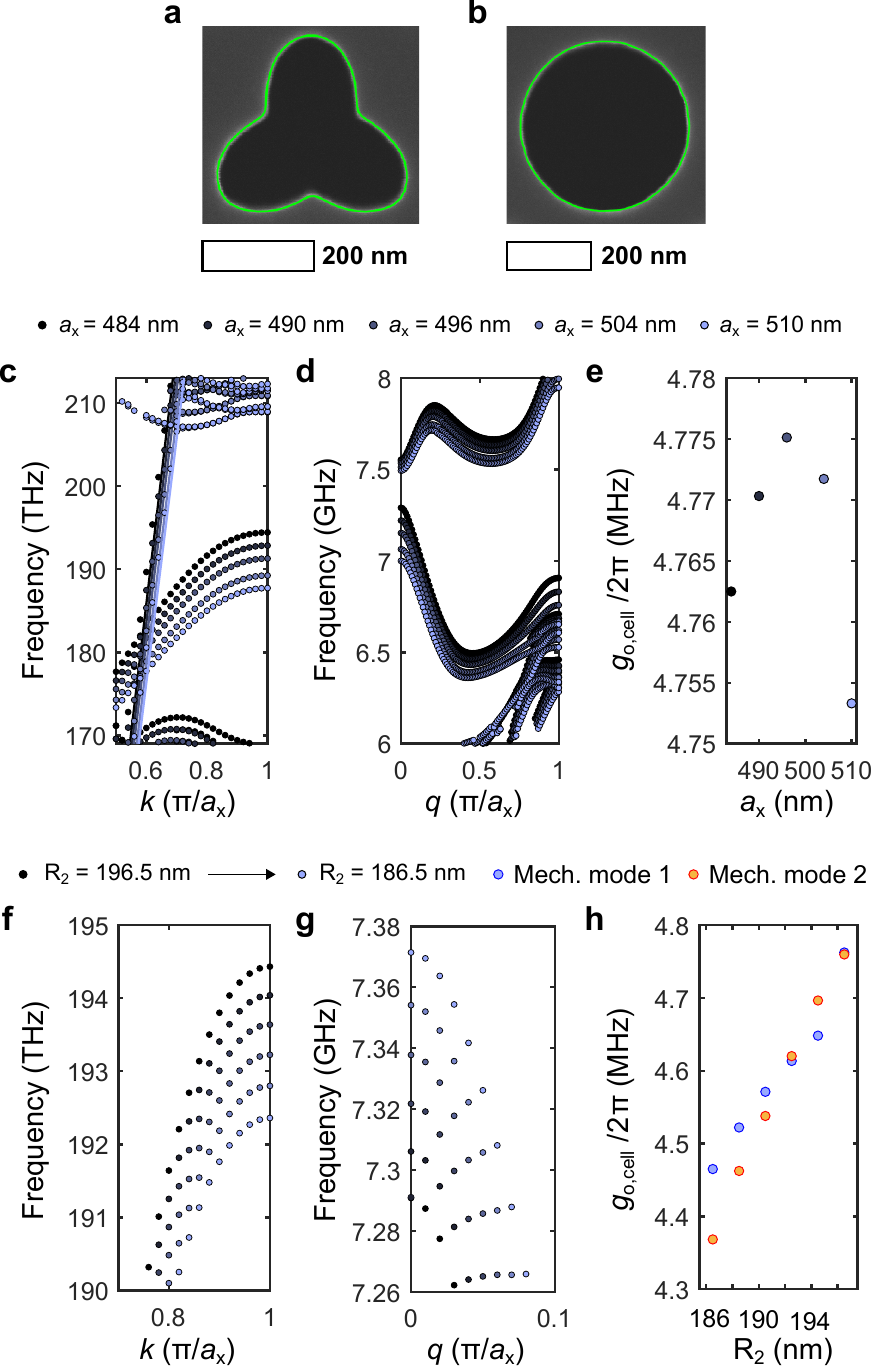}
\caption{\textbf{Optical and mechanical dispersion diagrams of the fabricated structures.} (a) Scanning electron microscopy (SEM) image of a representative shamrock-shaped hole and extracted average outline (green line). (b) Same as (a) for a circular hole. (c) Optical and (d) mechanical band structures for the waveguides with $R_1 = R_2$, $a_y$ = 484 nm and varying $a_x$. (e) The coupling rate $g_{\text{o,cell}}$ between the Bloch modes respectively at $q = 0$ and $k = \pi/a_x$ for the varying $a_x$. (f-h) Same as (c-e) for waveguides with $a_x$ = $a_y$ = 484 nm, $R_1$ = 196.5 nm and varying $R_2$.}
\label{fig:bands}
\end{figure}

In the waveguides we explore, shown in Fig. 1(a) in the main text, we vary several geometrical parameters. The most relevant ones are the horizontal pitch, $a_x$, and the radius of the circular hole in one of the membrane sides, $R_2$. The rest of the parameters are fixed, such as the vertical pitch, $a_y$ = 484 nm, or the different parameters defining the shamrock-shaped holes. We reconstruct the underlying band structures of the relevant waveguide unit cells (both for the waveguides themselves and for the heterostructure cavities) by extracting the contour of the fixed features from analysis of scanning electron microscopy (SEM) images. Figures~\ref{fig:bands}(a) and (b) show a high-magnification SEM image of a shamrock-shaped hole and a circular hole along with the outline (in green) extracted after averaging that of many holes of the same shape. The geometry of the shamrock is kept as extracted, while the circle is found to have a radius $R_{\text{1,fab}}$ = 196.5 nm. This evidences a slight overgrowth of the exposed and etched areas compared to the lithography mask ($R_{\text{1,mask}}$ = 194 nm). The fabricated samples also include varying slot widths, $s$, but we mainly focus here on MOM devices with extracted $s$ of 50 nm. Using the extracted geometric features and the centroid positions given by the triangular lattice parameters ($a_x$ and $a_y$), we calculate the dispersion diagrams using a commercial finite-element solver (COMSOL Multiphysics). Note that for the mechanics, we consider the anisotropy of the silicon stiffness tensor and its particular orientation with respect to the waveguide coordinate system~\cite{florez_engineering_2022,kersul_silicon_2023}. Figures~\ref{fig:bands}(c) and (d) show the optical and mechanical band diagrams for the case where $R_1 = R_2$, $a_y$ = 484 nm and varying $a_x$. We see that, as expected, the frequencies of the optical and mechanical bands of interest decrease as $a_x$ grows, which we use to build a rectangular confinement potential in the mirror-terminated long waveguides or a smooth potential well in the mode-gap adiabatic heterostructure cavities. We also provide in Fig.~\ref{fig:bands}(e) the unit cell vacuum optomechanical coupling rate, $g_{\text{o,cell}}/2\pi$, between the Bloch modes respectively at $q = 0$ and $k = \pi/a_x$ for the varying $a_x$. We observe nearly no variation of $g_{\text{o,cell}}/2\pi$ as a moving boundary contribution dominates it. The latter is mainly determined by the slot width, $s$, which is kept fixed. 

Figures~\ref{fig:bands}(f) and (g) depict the effect of reducing $R_2$ on the optical and mechanical band structure for the case $a_x$ = $a_y$ = 484 nm. We see that the reduced air-filling fraction as $R_2$ is reduced relative to $R_1$, which lowers the effective refractive index of the mode, leads to a redshift of the optical band. However, we note that the band edge of the mode is always well within the telecom C-band. In the case of the mechanical band structures, $\Delta R = R_1 - R_2 \neq 0$, breaks the $y$-symmetry and lifts the degeneracy between the frequencies of the modes on the two membrane sides. While the bands on one side are independent of $R_2$, the frequency of the mechanical mode on the other membrane side increases as $R_2$ is reduced. In the following section, we see how this effect is used in the case of the MOM system to control the frequency difference between the two cavity phonon modes.\\

\section{Two-dimensional optomechanical crystal cavities: cavity details}

\begin{figure}[ht]
\centering
\includegraphics[width=\linewidth]{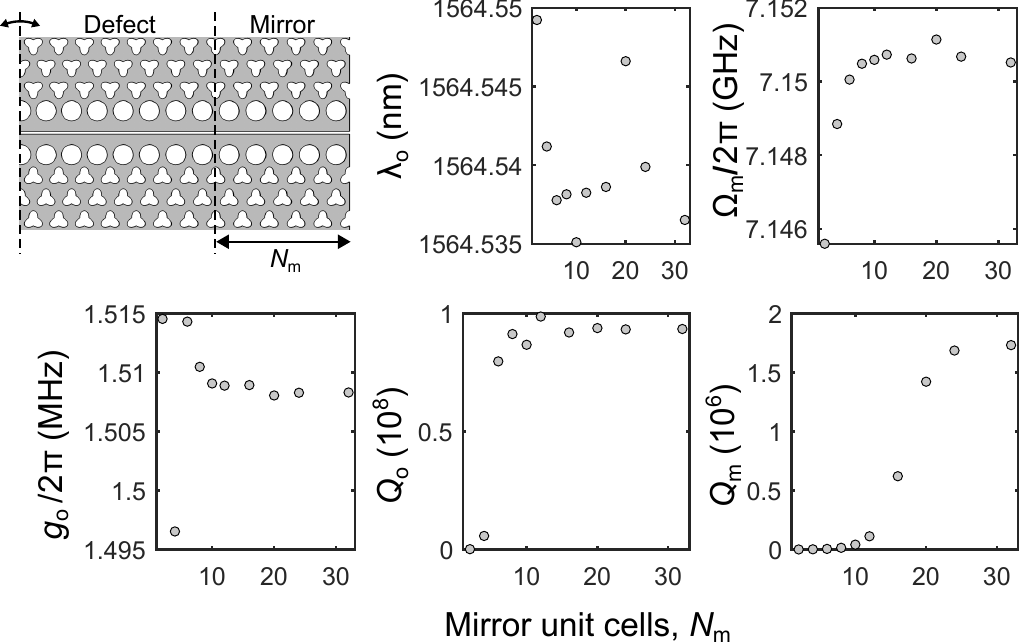}
\caption{\textbf{Cavity-optomechanical figures of merit as a function of the number of mirror unit cells.} Vacuum optomechanical coupling rates, $g_{\text{o}}$, optical wavelength, $\lambda_{\text{o}}$, optical quality factor, $Q_{\text{o}}$, mechanical frequencies, $\Omega_{\text{m}}$, and mechanical quality factors, $Q_{\text{m}}$ as a function of $N_m$. The mechanical and optomechanical figures of merit are the same for both degenerate mechanical modes.}
\label{fig:Nm_dependence}
\end{figure}

The mode-gap adiabatic heterostructure cavities are built by leveraging the band-structure properties depicted in Figs.~\ref{fig:bands}(c) and (d). The horizontal pitch of the waveguide, $a_x$, is changed adiabatically from $a_1$ = 484 nm in the central unit cell to $a_{N_c}$ = 510 nm at the edges of the defect region, which comprises $N_c = 8$ unit cells, following a cubic interpolation

\begin{equation*}
    a_i =  \Bigg\lfloor a_{N_c}\Bigg(1-\Big(1-\frac{a_1}{a_{N_c}}\Big)\Big(2\big(\frac{i}{N_c}\big)^3 - 3\big(\frac{i}{N_c}\big)^2 + 1\Big)\Bigg)\Bigg\rceil
\end{equation*}

where $i \in [0,N_c]$ and the pitch $a_i$ is rounded to integer values to conform with requirements for the lithographic mask/process. Such adiabatic tapering of $a_x$ is fixed for all the cavities we explore. Beyond the defect region, two mirror sections comprised of $N_{\text{m}}$  unit cells with pitch $a_{N_c}$ are included. To investigate the intrinsic frequencies and losses of both the optical and mechanical cavity modes, we explore the behaviour of the different cavity parameters as a function of $N_{\text{m}}$ (Fig.~\ref{fig:Nm_dependence}). Note that the simulated cavities use the geometry of the electron-beam lithography mask instead of the outlines of the fabricated holes, as in Fig~\ref{fig:bands}, because the smaller mesh elements required to capture the SEM-extracted contours would otherwise lead to prohibitive computational memory requirements. We see that nearly all cavity-optomechanical figures of merit except for the radiation-limited mechanical quality factor, $Q_{\text{m}}$, are converged ---with some numerical fluctuations but no trend--- for $N_{\text{m}}>10$. For the saturation of $Q_{\text{m}}$, 32 mirror unit cells are necessary. This large number is linked to the low effective mass associated with the dispersion at the band edge, which results in a low mirror strength~\cite{faggiani_lower_2016}. In addition, we observe that the value of $Q_{\text{m}}$ saturates, which implies that the size of the phononic-crystal cladding transverse to the waveguide axis is the ultimate limiting factor in our simulation setting. Our fabricated cavities use a cladding twice as wide as the simulations; therefore, the radiation-limited $Q_{\text{m}}$, i.e. at very low temperatures, can be expected to be much larger.

\begin{figure}[ht]
\centering
\includegraphics[scale=0.5]{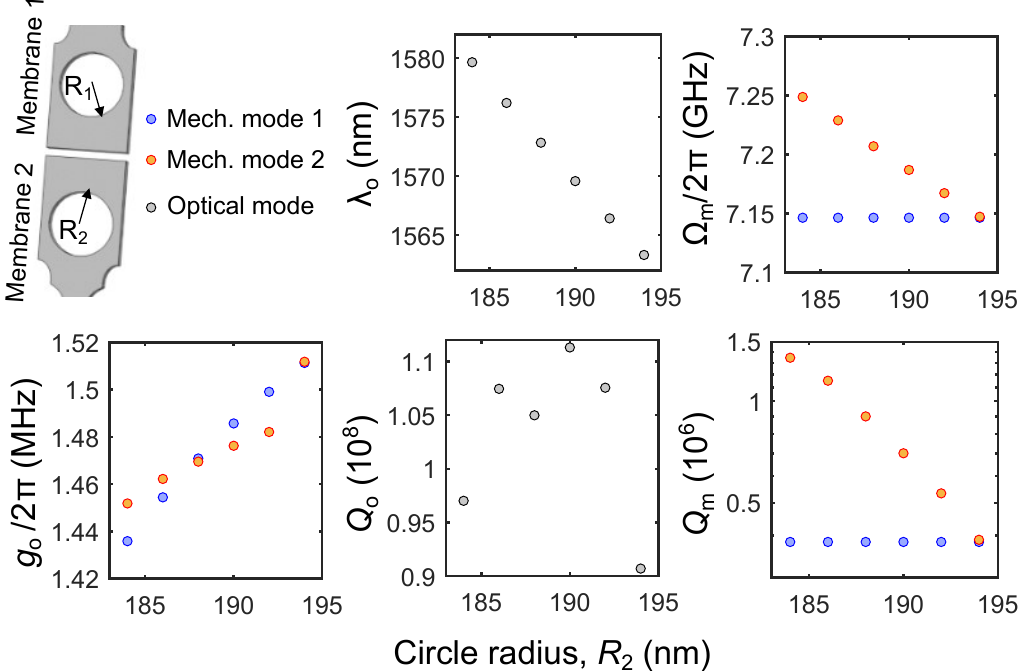}
\caption{\textbf{Cavity-optomechanical figures of merit as a function of the radius of the circular holes in the bottom membrane, $R_2$.} Vacuum optomechanical coupling rates, $g_{\text{o}}$, optical wavelength, $\lambda_{\text{o}}$, optical quality factor, $Q_{\text{o}}$, mechanical frequencies, $\Omega_{\text{m}}$, and mechanical quality factors, $Q_{\text{m}}$ as a function of $R_2$.}
\label{fig:r2_dependence}
\end{figure}

Similar to Fig.~\ref{fig:Nm_dependence}, we investigate the role of varying $R_2$ on the MOM system. In this case, the two mechanical modes are not degenerate anymore. The resulting cavity-optomechanical figures of merit for the case $N_{\text{m}}$ = 16 are shown in Fig.~\ref{fig:r2_dependence}. We observe the following: 1) the coupling rates, $g_{\text{o,1}}$ and $g_{\text{o,2}}$, which only coincide for the case $R_1$ = $R_2$ = 194 nm, depend very weakly on the value of $R_2$; 2) the resonant wavelength $\lambda_{\text{o}}$ redshifts by 1.63 nm per 1 nm decrease in $R_2$; 3) the optical $Q$ is not clearly affected by $R_2$; 4) the mechanical frequency $\Omega_{\text{m,2}}/2\pi$ increases by 10.19 MHz per 1 nm decrease in $R_2$; and 5) the mechanical quality factor grows steadily as $R_2$ increases. In connection with the discussion of the dependence of $Q_{\text{m}}$ with $N_{\text{m}}$, we attribute the behaviour with $R_2$ to the increase of the effective mass as $R_2$ decreases (see Fig.~\ref{fig:bands}(g)).\\

\section{Group velocity reconstruction}

\begin{figure}[ht]
\centering
\includegraphics[width=0.95\linewidth]{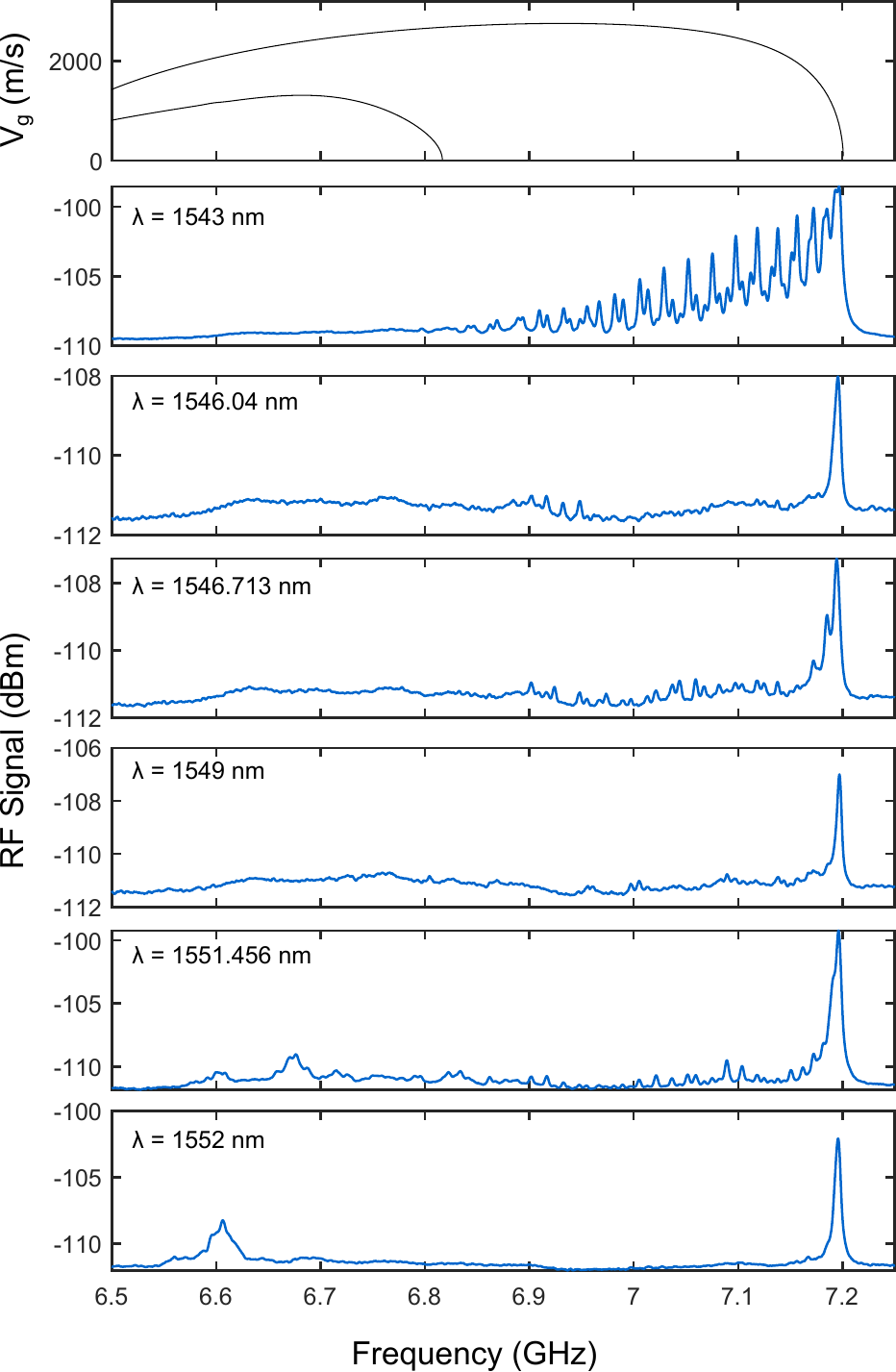}
\caption{\textbf{Radiofrequency spectra measured using different Anderson-localized modes on a waveguide made of 350 unit cells.} For reference, the simulated group velocity, $v_{\text{g}}$, shifted down by 90 MHz, is included as the top panel .}
\label{fig:RFSpectra}
\end{figure}

In Fig. 1(d) in the main text, we compare the simulated group velocity, $v_{\text{g}}$, of the mechanical mode of interest with that extracted via the free spectral range (FSR) of the transduced peaks in a waveguide of length $L$ = 350$a_x$. Although a single radiofrequency (RF) spectrum is provided in the main text, the reconstruction uses more RF spectra---albeit with poorer transduction amplitudes---to either confirm the modes observed in the spectrum given in the main text or to detect the presence of additional peaks that are difficult to identify in that single RF spectrum. Figure~\ref{fig:RFSpectra} reproduces the spectra used and also evidences that in the region where the acoustic waveguide is multimoded, the peaks can not be easily resolved, and there is no well-defined FSR~\cite{patel_single-mode_2018}. The persistent presence of the peak at the highest frequency might indicate that it is a tightly-localized acoustic mode, i.e. an Anderson-localized acoustic mode, but confirming such a hypothesis is beyond the scope of this work.

\section{Measurement of the vacuum optomechanical coupling rate}

We measure the vacuum optomechanical coupling rate, $g_0$, of the two mechanical modes of the MOM system following the method of Gorodetksy \textit{et al.}~\cite{gorodetksy_determination_2010}. The thermally excited displacement of the two mechanical modes is probed. Their power spectral density is compared to that transduced by the optical cavity for a phase-modulation tone induced by an electro-optic modulator (EOM) driven by a vector network analyzer (VNA) at $\omega_\text{VNA} = \SI{7.01}{GHz}$ and fixed power $\SI{-10}{dBm}$. This power corresponds to a modulation voltage $V_\text{VNA}=\SI{0.0707}{V}$ relative to the voltage required for a full phase-shift, i.e., $V_\pi(\SI{7}{GHz})=\SI{5.29}{V}$. The corresponding modulation depth is $d=\pi\frac{V_\mathrm{VNA}}{V_\pi}$ The two vacuum optomechanical coupling rates, $g_{o,1}$ and $g_{o,2}$, are evaluated as, \cite{gorodetksy_determination_2010}

\begin{figure}[ht]
\centering
\includegraphics[width=\linewidth]{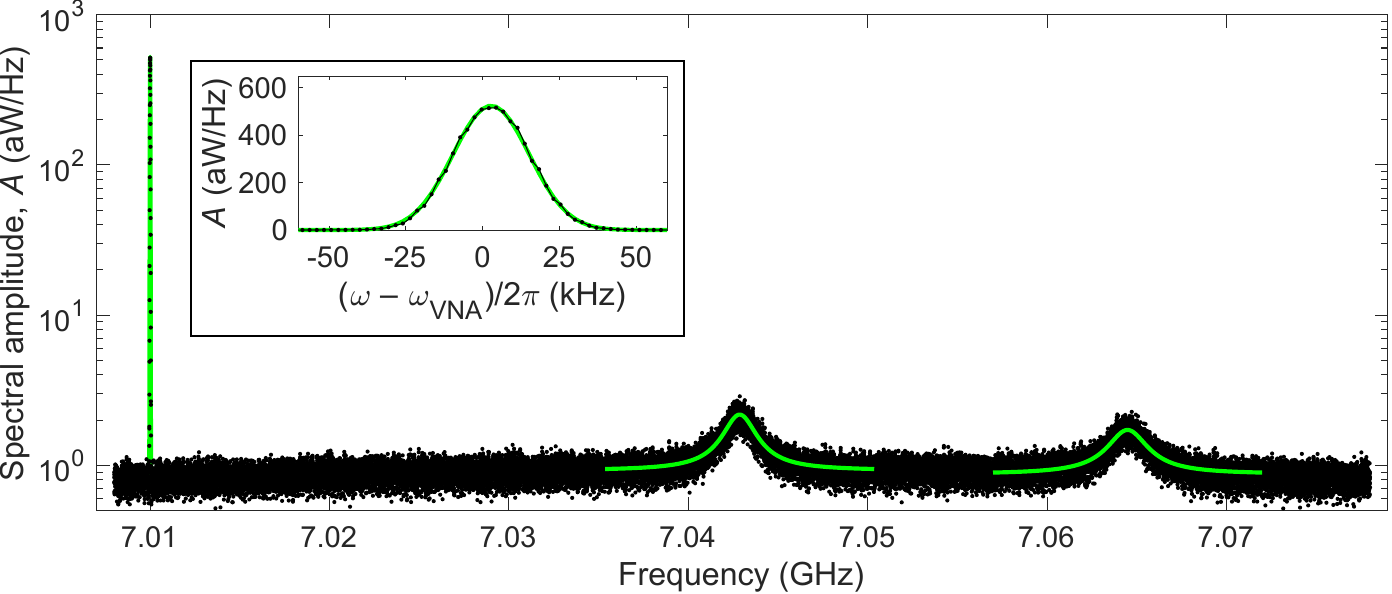}
\caption{
\textbf{Measurement of the vacuum optomechanical coupling,} $\mathbf{g_0}$.
The green lines represent Gaussian (Lorentzian) fits to the VNA (opto-mechanical) mode(s). The inset shows a magnification of the phase-modulation peak and the corresponding fit to a Gaussian.}
\label{fig:g0}
\end{figure}

\begin{equation}
\label{eq:go}
    \frac{g_{o,i}}{2\pi} = \frac{d\times\omega_\mathrm{VNA}}{4\sqrt{n}} \sqrt{\frac{A(\Omega_\mathrm{m,i})\Gamma_\mathrm{m,i}}{A(\omega_\mathrm{VNA})\gamma_\mathrm{mod}}}.
\end{equation}
Here, $A(\Omega_\mathrm{m,i})$ and $A(\omega_\mathrm{VNA})$ are the measured amplitudes of the power spectral density at the mechanical peak and at the modulation peak, respectively. $\Gamma_\mathrm{m,i}$ and $\gamma_\mathrm{mod}$ are the mechanical and modulation peak widths, repectively provided by the Lorentzian and Gaussian fits. $n=k_\text{B}T / (h \Omega_i)$ is the phonon thermal occupancy, with $k_\text{B}$, $T$ and $h$ the Boltzmann constant, the temperature and Planck's constant, respectively. Figure~\ref{fig:g0} shows the measured power spectral density, as well as two independent Lorentzian fits to the mechanical modes and a Gaussian fit to the VNA tone for a MOM system with $R_2$ = 192 nm. Table~\ref{table:g0} summarizes the results obtained from the fits and the evaluation of Eq.(~\ref{eq:go}).

\begin{table}[!ht]
\centering
\caption{Frequency ($\Omega$), linewidth ($\Gamma$), quality factor ($Q=\Omega/\Gamma$) and vacuum optomechanical coupling ($g_{\text{o}}/(2\pi)$) for the two mechanical modes.}
\begin{tabular}{ccccc}
\hline
Mode & $\Omega/2\pi$ (GHz) & $\Gamma/2\pi$ (MHz) & $Q$ & $g_0/2\pi$ (MHz) \\
\hline
1 (left) & \SI{7.043}{GHz} & \SI{2.04}{MHz} & $3464$ & \SI{1.543}{MHz}$\pm$ \SI{39}{kHz} \\
2 (right) & \SI{7.065}{GHz} & \SI{2.42}{MHz} & $2969$ & \SI{1.373}{MHz}$\pm$ \SI{45}{kHz} \\
\hline
\end{tabular}
  \label{table:g0}
\end{table}

\section{Effective laser detuning in a thermo-optic cavity}

Due to the thermo-optic nonlinearity, the laser wavelength $\lambda$ does not straightforwardly control the laser detuning to the cavity $\Delta$. Within a linear case, the optical cavity can be modelled using coupled mode theory and the transmission through the fiber loop is given by,
\begin{equation}
\label{transmission}
    T = \frac{\Delta^2+\kappa_i^2/4}{\Delta^2+\kappa_t^2/4}
\end{equation}
with $\kappa_i$ and $\kappa_e$ the internal and external decay rates and $\kappa_t=\kappa_i+\kappa_e$. The laser-cavity detuning, $\Delta$, is given by $\Delta=\omega-\omega_c$, with $\omega$ the laser frequency and $\omega_c$ the cavity resonance frequency. In the presence of a thermo-optic nonlinearity, $\beta$, the former is power-dependent and writes $\omega_c=\omega_0+\beta E$ with $E$ the intracavity energy and $\omega_0$ the cavity natural frequency (at zero input power). The value of $\beta$ depends on several optical and thermal properties of the host material and is also affected by the optical field profile~\cite{PhysRevB.102.245404}. The independent extraction of such parameters is cumbersome and subject to numerous sources of uncertainty.
\begin{figure}[!ht]
\centering
\includegraphics[width=\linewidth]{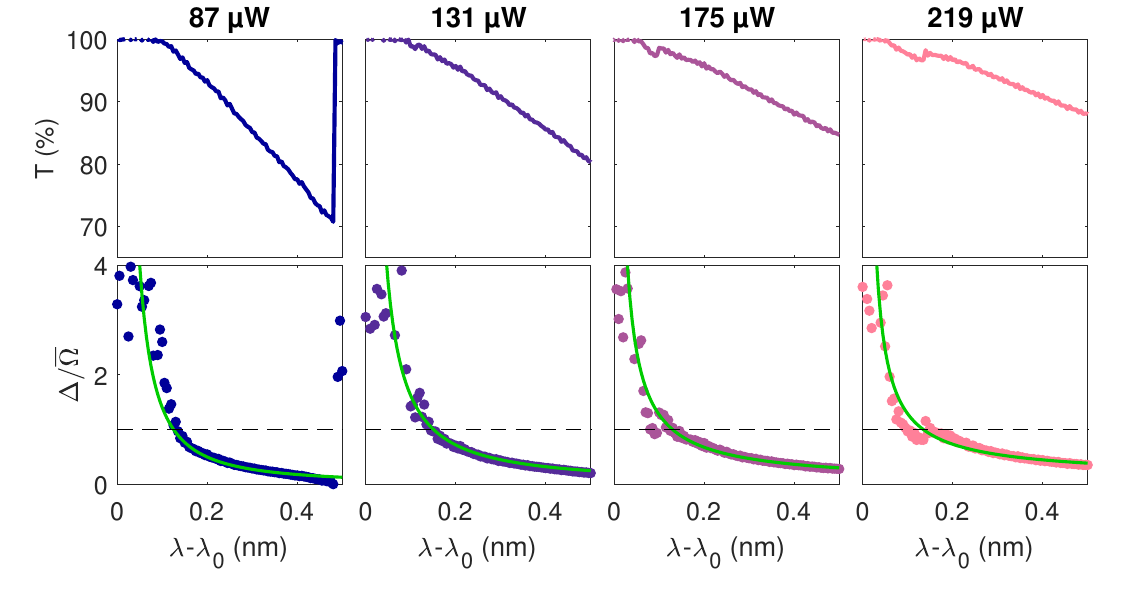}
\caption{\textbf{Extraction of the laser-cavity detuning as a function of the laser wavelength.} For increasing optical power launched to the cavity, we measure the normalized transmission (top) from which we infer the laser detuning $\Delta$ as a function of the laser wavelength $\lambda$. Blue-sideband excitation occurs at $\Delta=+\overline{\Omega}$, as indicated with the dashed lines. The evaluated detuning is fitted with a power law (green lines).}
\label{fig:asympt_fit}
\end{figure}
On the contrary, it is possible to evaluate the detuning from the measured transmission $T$ using Eq.~\ref{transmission},
\begin{equation}
\label{detuning}
    |\Delta| = \sqrt{\frac{1}{4}\frac{\kappa_t^2T-\kappa_i^2}{1-T}}
\end{equation}
This enables us to obtain the detuning as a function of, e.g., the laser wavelength $\lambda$ and establish a correspondence between the control parameter ($\lambda$) and the physically relevant quantity ($\Delta$). For several measurements of the transmission spectrum $T(\lambda)$ at increasing laser power (see Fig.~\ref{fig:asympt_fit}, top), we show the corresponding laser detuning $\Delta$ deduced from Eq.~\ref{detuning} (bottom). The data are plotted as a function of $\lambda-\lambda_0$ with $\lambda_0=2\pi c/\omega_0$ the natural cavity resonance wavelength evaluated to be $\lambda_0\sim1559.35$ nm. In absence of thermo-optic nonlinearity, the detuning would scale linearly with the laser wavelength in the case of the nearly-resonant excitations used here: $\Delta\approx\frac{2\pi c(\lambda_0-\lambda)}{\lambda_0^2}$. Instead, we observe with this representation how $\Delta$ scales asymptotically with $\lambda$. We fit the data with a power law $\Delta(\lambda) = A + B(\lambda-\lambda_0)^C$ (green lines). The fits are rather satisfying which justifies the use of semi-logarithmic scales in Fig. 3(c) of the main text for comparing the experiment to theory.

This method assumes that the cavity decay rates remain unchanged with the input power, which remains an approximation due to the emergence of two-photon absorption at higher power that tends to increase the intrinsic loss rate $\kappa_i$. However, we note an irregularity observed in the transmission curves that becomes stronger when the power is increased. This ``kink'' typically indicates optomechanical self-oscillations, which occurs when the laser reaches the blue sideband at $\Delta=+\overline{\Omega}$, where $\overline{\Omega}=\frac{1}{2}(\Omega_1+\Omega_2)$. Here, it appears in agreement with the determined laser detuning as shown with the dashed lines at $\Delta/\overline{\Omega}\approx+1$, where we use $\overline{\Omega}/2\pi=7.060$ GHz. It is also in agreement with the measurements presented in Fig.3d-e of the manuscript, where at least one mechanical mode becomes very close to its lasing threshold  $\Gamma_i=0$ for input powers above 131 $\mu$W. 

\section{Linearized optomechanical susceptibility}

We consider a Mechanical-Optical-Mechanical configuration (MOM), i.e. two mechanically independent mechanical modes with displacements $x_1$ and $x_2$, coupled to a single optical mode with amplitude $\ac$ via their respective optomechanical coupling rates $\go$ and $\ggo$. The classical equations of motion of this system are given by,

\begin{equation}
\label{eq1}
    \begin{aligned}
        \dot{\ac} &= \Big(j\big(\Delta + \go x_1+ \ggo x_2\big) - \frac{\kappa_t}{2}\Big)\ac + \sqrt{\frac{\kappa_e}{2}}\ain \\
        \ddot{x_1} &= -\Gamma_\mathrm{m,1} \dot{x_1} -\Omega_\mathrm{m,1}^2 x_1 + 2\Omega_\mathrm{m,1}\go |\ac|^2 \\ 
        \ddot{x_2} &= -\Gamma_\mathrm{m,2} \dot{x_2} -\Omega_\mathrm{m,2}^2 x_2 + 2\Omega_\mathrm{m,2}\ggo |\ac|^2
        \end{aligned}
\end{equation}
where $\Delta=\omega_\ell-\omega_0$ is the detuning between the laser frequency $\omega_\ell=2\pi c/\lambda$ and the cavity resonance frequency $\omega_0$, and $\ain$ is the input laser field amplitude, such that $\pin=\hbar\omega_\ell|\ain|^2$ is the laser incident power and $\ncav=|\ac|^2$ the cavity photon number. Note that the displacements are normalized by the respective zero-point fluctuations of the mechanical modes such that the mechanical masses can be eliminated.
 
Each mechanical resonator has a natural mechanical susceptibility $\chi_\mathrm{m,i}^{-1}(\omega) = \Big[ (\Omega_\mathrm{m,i}^2 - \omega^2) - j\omega\Gamma_\mathrm{m,i}\Big]$, that is perturbed by radiation pressure forces. The perturbation is described by a self-coupling term \cite{aspelmeyer2014cavity} $\Sigma_i(\omega) = 2\Omega_\mathrm{m,i}g_\mathrm{o,i}^2\beta(\omega)$ where $\beta(\omega) = \frac{\ncav}{(\Delta+\omega)+j\kappa_t/2}+\frac{\ncav}{(\Delta-\omega)-j\kappa_t/2}$.
The presence of a second mechanical mode leads to a deviation in the effective susceptibility of each mechanical oscillator. Applying a linearization to Eq.~\ref{eq1}, we obtain:
\begin{equation}
\label{eq2}
    \begin{aligned}
        \big[\chi_\mathrm{eff,1}^{-1}(\omega) + \Sigma_1(\omega)\big]x_1 = 2\Omega_\mathrm{m,1}\go\ggo\beta(\omega)x_2\\
        \big[\chi_\mathrm{eff,2}^{-1}(\omega) + \Sigma_2(\omega)\big]x_2 = 2\Omega_\mathrm{m,2}\go\ggo\beta(\omega)x_1
        \end{aligned}
\end{equation}
This expression highlights that the coupling between the mechanical modes scales with $g_1g_2$, which is expected since they do not interact directly but through the optical field. Note that taking $\Gamma_1\approx\Gamma_2$ provides the Hamiltonian formulation presented in Ref.~\cite{xu2016topological}.
Solving Eq.~\ref{eq2} provides an expression for the effective mechanical coupling between the mechanical modes:

\begin{equation}
\label{K_eq}
K_i(\omega) = -\frac{\Sigma_i(\omega)\Sigma_j(\omega)}{\chi_\mathrm{m,j}^{-1}(\omega) + \Sigma_j(\omega)}
\end{equation}

\begin{figure}[!ht]
    \centering
   \includegraphics[scale=0.7,trim={0.8cm 0.2cm 0.9cm 0.2cm},clip]{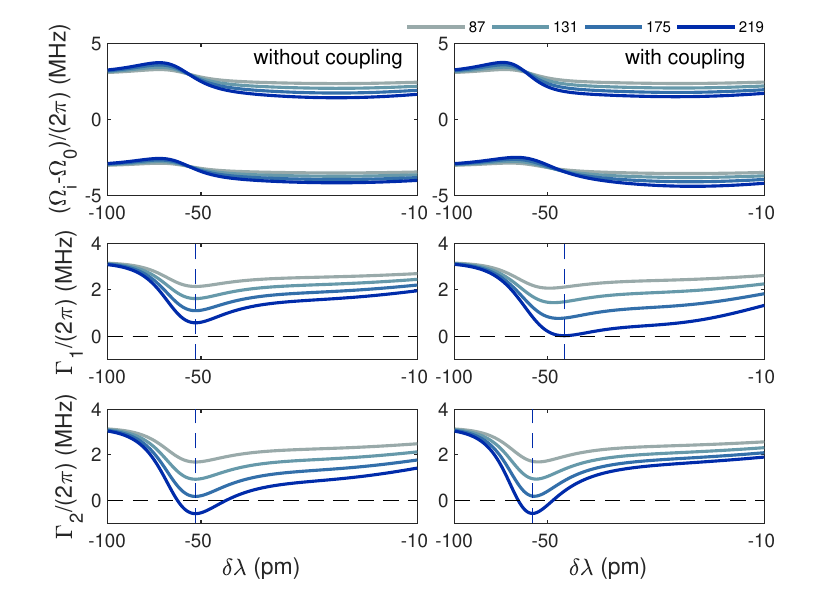}
    \caption{\textbf{The role of the optically-induced mechanical coupling in the observed dynamical back-action effects.} Comparison of the theory curves presented in Fig.~3 of the main text when accounting for a MOM with effective coupling (right) compared to a MOM with no optically-mediated coupling between the two mechanical modes (left). The mechanical frequencies and linewidths are shown for different input laser power indicated in units of $\mu$W.}
    \label{fig:nocoupling}
\end{figure}

Therefore we can write the total mechanical susceptibility of mode \textit{i} as follow:

\begin{equation}
\label{totSusc}
    \chi_\mathrm{eff,i}^{-1}(\omega) = \chi_\mathrm{m,i}^{-1}(\omega) +\Sigma_i(\omega)+ K_i(\omega)
\end{equation}

From Eq.~\ref{totSusc}, we evaluate the effective mechanical frequencies and damping rates using $\Omega_\mathrm{i}=\Omega_\mathrm{m,i}+\mathrm{Re}[\Sigma_i(\omega)+ K_i(\omega)]/(2\Omega_i)$, and $\Gamma_\mathrm{i}=-\mathrm{Im}[\chi_\mathrm{eff,i}^{-1}]/\Omega_i$, respectively. These are the theoretical quantities plotted as a function of $\lambda$ and $\pin$ in Fig.~3 in the main text.

In Fig.~\ref{fig:nocoupling}, we compare the theoretical evolutions of $\Omega_1$, $\Omega_2$, $\Gamma_1$, and $\Omega_2$ as a function of the detuning ($\delta\lambda$, expressed in terms of wavelength). We compare the situation where the mechanical oscillator do not interact through the optical mode (i.e., by imposing $K_i=0$) with the MOM configuration (with $K_i$ given by Eq.~\ref{K_eq}). We note a significant change in the evolution of the linewidths. In particular, accounting for the coupling enables the recovery of two features: 1) the mode linewidths reach a minimum at different spectral locations (indicated by dashed lines) which is not the case when considering single optomechanical cavities, and 2) the decrease of $\Gamma_1$ is more pronounced, enabling the mode to pass its lasing threshold ($\Gamma_1=0$) at the power $\pin=219$ $\mu$W, while the linewidth of the other mode reduces as much, but within a shorter spectral range. In summary, in the absence of coupling the frequency and linewidth of mode 1 and 2 evolve identically, with only a scaling factor and an offset. However accounting for the full MOM mechanical susceptibility reveals that the modes follow different spectral evolutions (i.e., along the $x$-axis). 

\section{Stimulated multimode lasing: methodology}

Here we provide more details on the theoretical curves included in Figure 4(c) in the main text. We present, first, the methodology used to determine the lasing threshold from the measurement of the mechanical peak amplitudes as a function of the modulation depth, $d$, and frequency, $\omega_\mathrm{VNA}$. Secondly, we detail how the extracted thresholds are fitted using the laser detuning $\Delta$ as a fitting parameter and the theory described in Ref.~\cite{mercade_floquet_2021}.

\begin{figure}[ht]
\centering
\includegraphics[scale=0.55,trim={1cm 0.2cm 1.5cm 0.5cm},clip]{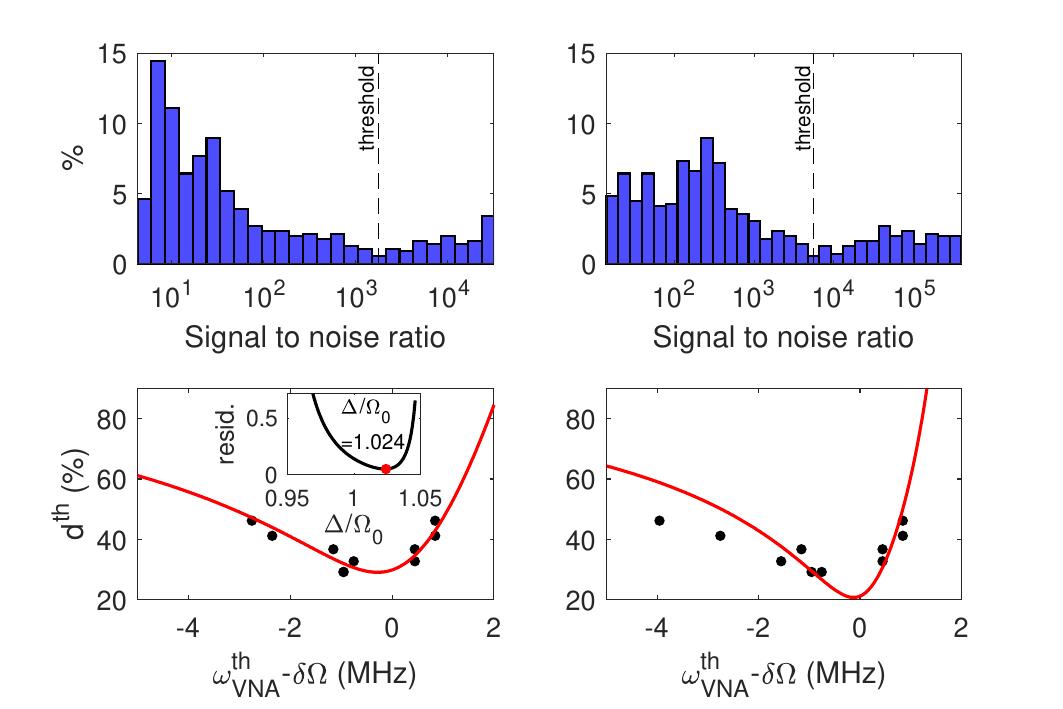}
\caption{\textbf{Lasing threshold determination and fitting} a. Distribution signal-to-noise ratio of mechanical modes 1 (left) and 2 (right) over the colormaps presented in Fig.~4 of the manuscript. The minimum of each distribution is interpreted as the lasing threshold amplitude. b. Lasing threshold of each mode (1:left and 2:right) plotted in the parameter space $\{ \omega_\mathrm{VNA}^\mathrm{th},d^\mathrm{th} \}$ (black dots) and simultaneously fitted with the theoretical model from Ref.\cite{mercade_floquet_2021} (red lines). The residuals (inset) yield a minimum at $\Delta/\overline{\Omega}=1.024$.}
\label{fig:tongue_fitting}
\end{figure}

The amplitude modulation of the input laser leads to sidebands of the mechanical modes in the radiofrequency spectrum~\cite{madiot2020coherent}. When $\delta\omega\equiv\omega_\mathrm{VNA}-(\Omega_2-\Omega_1)<\Gamma_{1,2}$, a sideband overlaps with the mechanical modes. This situation complicates the fitting of the mechanical modes with a Lorentzian lineshape, and the mechanical linewidth cannot be accurately determined. In order to determine the lasing thresholds -- given by $\Gamma_i=0$, by definition -- we use the fact that the peak amplitude grows quickly at the threshold. For each map presented in the Figure 4(c) in the main text, i.e., for each mechanical mode, we plot the statistics of the peak amplitude within the full map (see Fig.~\ref{fig:tongue_fitting}(a)). The statistics display a minimum (dashed lines) that we interpret as the value of the oscillator amplitude that indicates the threshold. For each mode, we then determine the positions in the parameter space, $\{ \omega_\mathrm{VNA}^\mathrm{th},d^\mathrm{th} \}$, where the amplitude passes this threshold. This provides the data points in Fig.~\ref{fig:tongue_fitting}(b). 

Now, we theoretically determine the threshold condition in the same parameter space. For a given mode \textit{i}, it corresponds to the ensemble of pairs $\{ \omega_\mathrm{VNA},d \}$ fulfilling the condition $\Gamma_i=0$. The model in Ref.~\cite{mercade_floquet_2021} leads to
\begin{equation*}
    \Gamma_i = -\frac{1}{\Omega_i}\mathrm{Im}\Big[\chi_\mathrm{m,i}^{-1} + \Sigma_i(\Omega_\mathrm{m,1}) - \frac{2\Omega_i\sigma^2(\Omega_i)}{\delta\omega + j\frac{\Gamma_\mathrm{m,i}}{2}+\sigma'(\Omega_i)}\Big]
\end{equation*}
with 
\begin{align*}
    \sigma(\omega)&=\go\ggo\ncav d \beta(\omega) \\
    \sigma'(\omega)&=\go\ggo\ncav d^2 \beta(\omega)
\end{align*}
and where $\chi_\mathrm{m,i}$, $\Sigma_i(\omega)$ and $\beta(\omega)$ are defined in the previous section,.

The fit is performed using $\Delta$ as the only fitting parameter, as it can be noted that both $\beta(\omega)$ and $\ncav$ depends on $\Delta$. The thresholds for mode 1 and 2 are fitted at once (see red line in Fig.~\ref{fig:tongue_fitting}(b)) and provides $\Delta=1.024\times\overline{\Omega}$ using the mean mechanical frequency $\overline{\Omega}=6.984$ GHz. The residual is shown in the inset.

\clearpage

\bibliography{MOM2}

\end{document}